\documentclass[notitlepage,aip]{revtex4-1}
\usepackage{graphicx}
\usepackage{multirow}
\usepackage{physics}
\usepackage{xcolor}
\usepackage[]{algorithm}

\draft 

\begin{document}

\author{D. Zhu \footnotemark[2]}

\affiliation{Joint  Quantum  Institute,  University  of  Maryland,  College  Park,  Maryland  20742,  USA}
\affiliation{Center for Quantum Information and Computer Science,
University of Maryland, College Park, MD 20742, USA}
\affiliation{Department of Electrical and Computer Engineering,  University  of  Maryland,  College  Park,  Maryland  20742,  USA}

\author{Z.P. Cian \footnote[1]{email : zpcian@umd.edu}\footnote[2]{These two authors contributed equally.}}

\affiliation{Joint  Quantum  Institute,  University  of  Maryland,  College  Park,  Maryland  20742,  USA}
\affiliation{Department  of  Physics,  University  of  Maryland,  College  Park,  Maryland  20742,  USA}

\author{C. Noel}
\affiliation{Joint  Quantum  Institute,  University  of  Maryland,  College  Park,  Maryland  20742,  USA}
\affiliation{Center for Quantum Information and Computer Science,
University of Maryland, College Park, MD 20742, USA}
\affiliation{Department  of  Physics,  University  of  Maryland,  College  Park,  Maryland  20742,  USA}
\affiliation{Duke Quantum Center and Department of Physics, Duke University, Durham, NC 27708,  USA}
\affiliation{Department of Electrical and Computer Engineering, Duke University, Durham, NC 27708,  USA}

\author{A. Risinger}
\affiliation{Joint  Quantum  Institute,  University  of  Maryland,  College  Park,  Maryland  20742,  USA}
\affiliation{Center for Quantum Information and Computer Science,
University of Maryland, College Park, MD 20742, USA}
\affiliation{Department of Electrical and Computer Engineering,  University  of  Maryland,  College  Park,  Maryland  20742,  USA}

\author{D. Biswas}
\affiliation{Joint  Quantum  Institute,  University  of  Maryland,  College  Park,  Maryland  20742,  USA}
\affiliation{Center for Quantum Information and Computer Science,
University of Maryland, College Park, MD 20742, USA}
\affiliation{Department  of  Physics,  University  of  Maryland,  College  Park,  Maryland  20742,  USA}

\author{L. Egan}
\affiliation{Joint  Quantum  Institute,  University  of  Maryland,  College  Park,  Maryland  20742,  USA}
\affiliation{Center for Quantum Information and Computer Science,
University of Maryland, College Park, MD 20742, USA}
\affiliation{Department  of  Physics,  University  of  Maryland,  College  Park,  Maryland  20742,  USA}

\author{Y. Zhu}
\affiliation{Joint  Quantum  Institute,  University  of  Maryland,  College  Park,  Maryland  20742,  USA}
\affiliation{Department  of  Physics,  University  of  Maryland,  College  Park,  Maryland  20742,  USA}

\author{A. M. Green}
\affiliation{Joint  Quantum  Institute,  University  of  Maryland,  College  Park,  Maryland  20742,  USA}
\affiliation{Department  of  Physics,  University  of  Maryland,  College  Park,  Maryland  20742,  USA}

\author{C. Huerta Alderete}
\affiliation{Joint  Quantum  Institute,  University  of  Maryland,  College  Park,  Maryland  20742,  USA}
\affiliation{Department  of  Physics,  University  of  Maryland,  College  Park,  Maryland  20742,  USA}

\author{N. H. Nguyen}
\affiliation{Joint  Quantum  Institute,  University  of  Maryland,  College  Park,  Maryland  20742,  USA}
\affiliation{Department  of  Physics,  University  of  Maryland,  College  Park,  Maryland  20742,  USA}

\author{Q. Wang}
\affiliation{Joint  Quantum  Institute,  University  of  Maryland,  College  Park,  Maryland  20742,  USA}
\affiliation{Center for Quantum Information and Computer Science,
University of Maryland, College Park, MD 20742, USA}
\affiliation{Department of Chemistry,
University of Maryland, College Park, MD 20742, USA}

\author{A. Maksymov}
\affiliation{IonQ,  College  Park,  Maryland  20740,  USA}

\author{Y. Nam}
\affiliation{IonQ,  College  Park,  Maryland  20740,  USA}
\affiliation{Department  of  Physics,  University  of  Maryland,  College  Park,  Maryland  20742,  USA}

\author{M. Cetina}
\affiliation{Joint  Quantum  Institute,  University  of  Maryland,  College  Park,  Maryland  20742,  USA}
\affiliation{Center for Quantum Information and Computer Science,
University of Maryland, College Park, MD 20742, USA}
\affiliation{Department  of  Physics,  University  of  Maryland,  College  Park,  Maryland  20742,  USA}
\affiliation{Duke Quantum Center and Department of Physics, Duke University, Durham, NC 27708,  USA}

\author{N. M. Linke}
\affiliation{Joint  Quantum  Institute,  University  of  Maryland,  College  Park,  Maryland  20742,  USA}
\affiliation{Department  of  Physics,  University  of  Maryland,  College  Park,  Maryland  20742,  USA}

\author{M. Hafezi}
\affiliation{Joint  Quantum  Institute,  University  of  Maryland,  College  Park,  Maryland  20742,  USA}
\affiliation{Department of Electrical and Computer Engineering,  University  of  Maryland,  College  Park,  Maryland  20742,  USA}
\affiliation{Department  of  Physics,  University  of  Maryland,  College  Park,  Maryland  20742,  USA}

\author{C. Monroe}
\affiliation{Joint  Quantum  Institute,  University  of  Maryland,  College  Park,  Maryland  20742,  USA}
\affiliation{Center for Quantum Information and Computer Science,
University of Maryland, College Park, MD 20742, USA}
\affiliation{Department  of  Physics,  University  of  Maryland,  College  Park,  Maryland  20742,  USA}
\affiliation{Duke Quantum Center and Department of Physics, Duke University, Durham, NC 27708,  USA}
\affiliation{Department of Electrical and Computer Engineering, Duke University, Durham, NC 27708,  USA}
\affiliation{IonQ,  College  Park,  Maryland  20740,  USA}



\title{Cross-Platform Comparison of Arbitrary Quantum Computations} 

\date{\today}

\pacs{}

\maketitle 
{\bf As we approach the era of quantum advantage, when quantum computers (QCs) can outperform any classical computer on particular tasks\cite{MikeAndIke}, there remains the difficult challenge of how to validate their performance. While algorithmic success can be easily verified in some instances such as number factoring \cite{shor1994algorithms} or oracular algorithms \cite{QuantumAlgorithmZoo}, these approaches only provide pass/fail information for a single QC. On the other hand, a comparison between different QCs on the same arbitrary circuit provides a lower-bound for generic validation: a quantum computation is only as valid as the agreement between the results produced on different QCs. Such an approach is also at the heart of evaluating metrological standards such as disparate atomic clocks \cite{Clocks2021}. In this paper, we report a cross-platform QC comparison using randomized and correlated measurements that results in a wealth of information on the QC systems. We execute several quantum circuits on widely different physical QC platforms and analyze the cross-platform fidelities.} 

Cross-platform quantum circuit comparisons are critical in the early stages of developing QC systems, as they may expose particular types of hardware-specific errors and also inform the fabrication of next-generation devices. There are straightforward methods for comparing generic output from different quantum computers, such as coherently swapping information between them \cite{buhrman2001quantum}, and full quantum state tomography \cite{Blume_Kohout_2010}. However, these schemes require either establishing a coherent quantum channel between the systems \cite{Kimble2008}, which may be impossible with highly disparate hardware types; or transforming quantum states to classical measurements, requiring resources that scale exponentially with system size. 
 
Recently, a new type of cross-platform comparison based on randomized measurements has been proposed \cite{crossplatfo,PRXQuantum.2.010102}.  
While this approach still scales exponentially with the number of qubits, it has a significantly smaller exponent prefactor compared with full quantum state tomography, allowing scaling to larger quantum computer systems. 

Here, we demonstrate a cross-platform comparison based on randomized-measurement \cite{crossplatfo, PRXQuantum.2.010102, huang2020predicting}, obtained independently over different times and locations on several disparate quantum computers built by different teams using different technologies, comparing the outcomes of four families of quantum circuits. We use four ion-trap platforms, the University of Maryland (UMD) EURIQA system \cite{egan2020fault} (referred to as UMD\_1), the University of Maryland TIQC system \cite{zhu2020generation} (UMD\_2), and two IonQ quantum computers \cite{Wright2019,Li2020} (IonQ\_1, IonQ\_2), as well as five separate IBM superconducting quantum computing systems hosted in New York, \emph{ibmq\_belem} (IBM\_1), \emph{ibmq\_casablanca} (IBM\_2), \emph{ibmq\_melbourne} (IBM\_3), \emph{ibmq\_quito} (IBM\_4), and \emph{ibmq\_rome} (IBM\_5) \cite{IBMquantum}. 
See Supplementary Information Sec.~\ref{supp:hardware} for more details of these systems.

We first demonstrate the application of randomized measurements for comparing  5-qubit GHZ (Greenberger–Horne–Zeilinger) states\cite{ghz} generated on different platforms and the ideal 5-qubit GHZ state obtained from classical simulation. Using the same protocol, we also compare states generated with three random circuits of different width and depth, each sharing a similar construction to circuits used in quantum volume (QV) measurements \cite{jurcevic2021demonstration}. 

\begin{figure}[h]
\centering
 \includegraphics[width=0.95\textwidth]{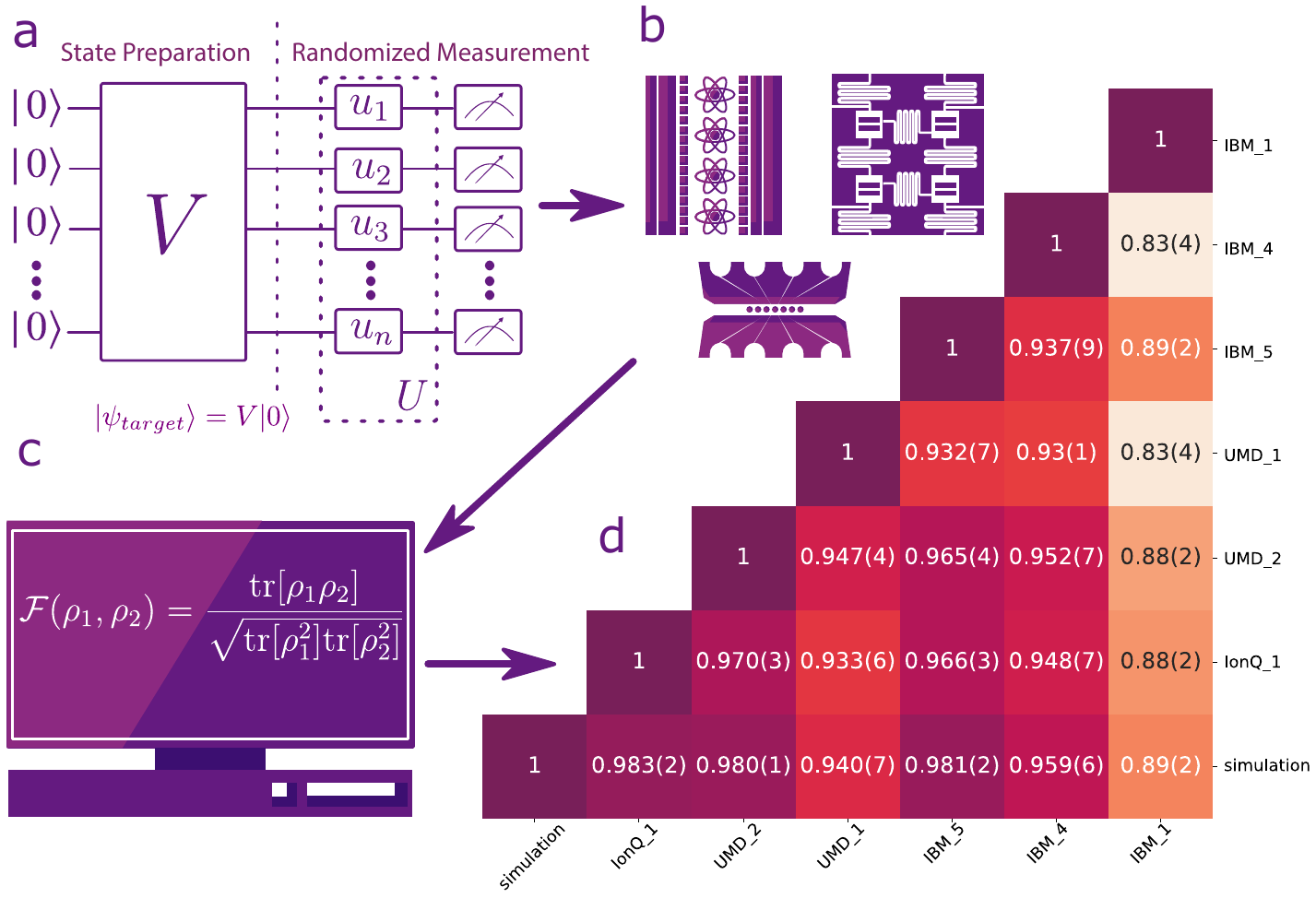}
 \caption{Schematic diagram of the cross-platform comparison. 
 {\bf a} Test quantum circuit, represented by unitary operator $V$ for state preparation, with appended random rotations $u_i$ to each qubit $i$ for measurements in a random (particular) basis.
 {\bf b} The circuits are transpiled for different quantum platforms into their corresponding native gates. Each of the $M_U$ circuits is repeated $M_S$ times for each platform. {\bf c} The measurement results are sent to a central data repository for processing the fidelities defined in Eq. \eqref{eq:x_platform_fidelity}. As an example, {\bf d} shows the cross-platform fidelity results for a 5-qubit GHZ state, including a row of comparisons between each of the six hardware systems and theory (labeled ``simulation").  Entry $i,j$ corresponds to the cross-platform fidelity between platform-$i$ and platform-$j$. The cross-platform fidelity is inferred from $M_U = 100$ randomized measurements and $M_S = 2000$ repetitions for each $U$.
 }
 \label{fig:x_platform_intro}
\end{figure}

The cross-platform fidelity that we use is defined as \cite{crossplatfo,liang2019quantum}
\begin{equation}
    \mathcal{F}(\rho_1, \rho_2) = \frac{{\rm tr}[\rho_1 \rho_2]}{\sqrt{{\rm tr}[\rho_1^2] {\rm tr}[\rho_2^2]}},\label{eq:x_platform_fidelity}
\end{equation}
where $\rho_i$ is the density matrix of the desired quantum state produced by system $i$. To evaluate this fidelity, for each system, we first initialize $N$ qubits in the state $|0,0,\dots, 0\rangle$ and apply the unitary $V$ to nominally prepare the desired quantum states on each platform. In order to measure the quantum states in $M_U$ different bases, we sample $M_U$ distinct combinations of random single-qubit rotations $ U = u_1 \otimes u_2 \otimes \dots \otimes u_N$ and append them to the circuit that implements $V$ as shown in Fig.~\ref{fig:x_platform_intro} a. Finally, we perform projective measurements in the computational basis. For each rotation setting $U$, the measurements are repeated $M_S$ times(``shots'') on each platform. 

The fidelity can be inferred from the randomized measurement results via either the statistical correlations between the randomized measurements\cite{crossplatfo} (Protocol I) or constructing an approximate classical representation of a quantum state using randomized measurements, the so-called the classical shadow \cite{huang2020predicting, aaronson2019shadow} (Protocol II).
In Protocol I, we calculate the second-order cross-correlations \cite{crossplatfo} between the outcomes of the two platforms $i$ and $j$ via the relation
\begin{equation}
    {\rm Tr}[\rho_i \rho_j] = 2^N \sum_{s, s'} (-2)^{-D[s, s']}\overline{P^{(i)}_U(s) P^{(j)}_U(s')},
    \label{eq:innsbruck}
\end{equation}
where $i,j \in \{ 1, 2\}$, $s = s_1, s_2,...,s_N$ is the bit string of the binary measurement outcomes $s_k$ of $k$th qubit, $D[s, s']$ is the Hamming distance between $s$ and $s'$, $P^{(i)}_U(s) = {\rm Tr}[U\rho_iU^\dagger |s \rangle \langle s |]$, and the overline 
denotes the average over random unitaries $U$.

For Protocol II, we reconstruct the classical shadow of the quantum state for each shot of measurement as $\hat{\rho} = \bigotimes_{k = 1}^N (3u_k^\dagger|s_k\rangle \langle s_k |u_k - I)$, where $I$ is the $2\times 2$ identity matrix \cite{aaronson2019shadow, huang2020predicting}. The overlap can be calculated as \cite{huang2020predicting}
\begin{equation}
    {\rm Tr}[\rho_i \rho_j] = \overline{{\rm Tr}[\hat{\rho}_i\hat{\rho}_j]},
    \label{eq:shadow}
\end{equation}
where $i,j \in \{ 1,2 \}$ and the overline denotes the average over all the experimental realizations. We note that, for both protocols, unbiased estimators are necessary when calculating the purity $i=j$\cite{huang2020predicting,crossplatfo} using Eq. \eqref{eq:innsbruck} and \eqref{eq:shadow}. 

While the fidelity inferred from the two protocols is identical in the asymptotic limit with $M=M_S \times M_U \rightarrow \infty$, the fidelity error inferred from Protocol II converges faster in the number of random unitaries \cite{huang2020predicting}. Therefore, we implement Protocol II for 5- and 7-qubit experiments. However, this protocol is more costly for post-processing. Therefore, for the 13-qubit experiment, we post-process the result with Protocol I.

We explore two different schemes for sampling the single-qubit unitary rotations $U$, a random method and a greedy method. In the regime $M_S \gg 2^N$, we observe that the greedy method outperforms the random method (see supplementary Information Sec. \ref{supp:greedy_random}). Therefore, for $N = 5,7$, we sample the single-qubit unitary operation with the greedy method. For $N=13$, we use the random method because to satisfy $M_S \gg 2^N$, the total number of measurements becomes too large. The specified target states and rotations are sent to each platform as shown in Fig.~\ref{fig:x_platform_intro}b,c. The circuit that implements the specified unitary $UV$ are synthesized and optimized for each platform in terms of its native gates.

When preparing a quantum state on a quantum system, one can perform various error-mitigation and circuit optimization techniques. While these techniques can greatly simplify the circuit and reduce the noise of the measurement outcomes, they can make the definition of state preparation ambiguous. For example, when we prepare a GHZ state and perform the projective measurement in the computational basis, we can defer the CNOT gates right before the measurement to the post-processing, instead of physically applying them. Although one can still obtain the same expectation value for any observable using such a circuit optimization technique, the GHZ state is not actually prepared in the quantum computer. In order to standardize the comparison, in this study, we require that one can perform arbitrary error-mitigation and circuit optimization techniques provided that the target state $|\psi_{target}\rangle = V|0\rangle$ is prepared at the end of the state-preparation stage.




After performing the experiments, the results are sent to a data repository. Finally, we process the results and calculate the cross-platform fidelities. 
The statistical uncertainty of the measured fidelity is inferred directly from the measurement results via a bootstrap resampling technique \cite{efron1983leisurely}. The bootstrap resampling allows us to evaluate the statistical fluctuation of the measurements as well as the system performance fluctuation within the duration of the data taking, which is typically two to three days. However, we note that it does not show system performance variations on longer time scale.




 We first measure the cross-platform fidelity to compare 5-qubit GHZ states. Specifically, the circuit that prepares the GHZ states are appended with a total of 243 different sets of single-qubit Clifford gates. Each appended circuit is repeated for $M_S = 2000$ shots. We sample $M_U=100$ out of the 243 different $U$s to calculate the cross-platform fidelity defined in Eq. \eqref{eq:x_platform_fidelity} (Fig.~\ref{fig:x_platform_intro}d). We see that our method has good enough resolution to reveal the performance difference between platforms. In Supplementary Information sec. \ref{supp:5q-ghz}, we benchmark our method against full quantum state tomography by computing the fidelity as a function of $M_U$. The comparison shows that the fidelity obtained via randomized measurements approaches that obtained via the full quantum state tomography rapidly. 

\label{sec:quantum-volumn-circuit}

\begin{figure}[h]
\centering
 \includegraphics[width=0.95\textwidth]{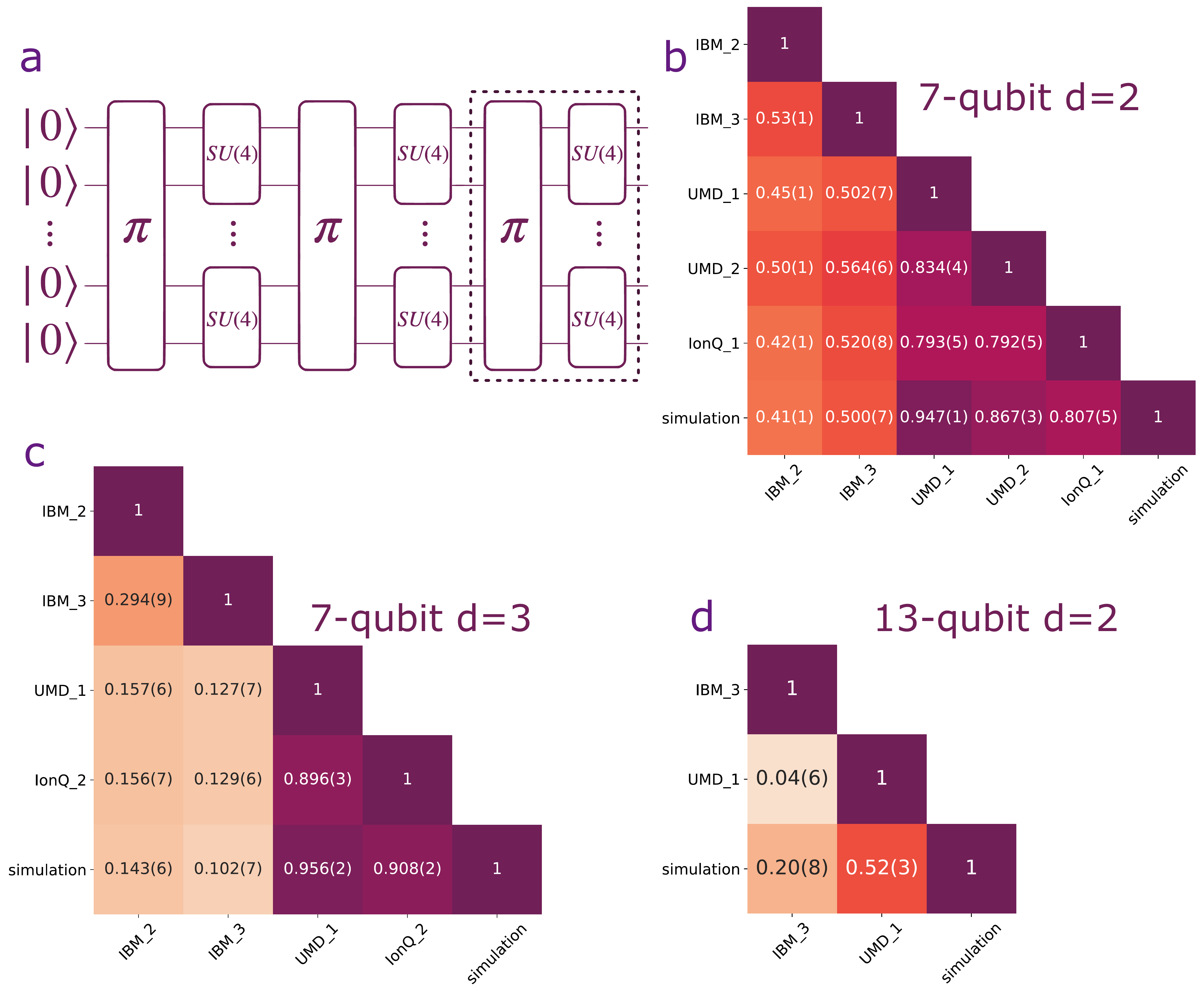}
 \caption{
 {\bf a} The quantum volume circuit diagram for $d=3$. The $d=2$ case does not have the operations in the dashed rectangle.  
     {\bf b} to {\bf d}  Cross-platform fidelity between different quantum computers. Entry $i,j$ corresponds to the cross-platform fidelity $\mathcal{F}(\rho_i, \rho_j)$ between platform-$i$ and platform-$j$ as defined in Eq.~\ref{eq:x_platform_fidelity}. 
    {\bf b} $N=7$ and $d=2$;
    {\bf c} $N=7$ and $d=3$;
    {\bf d} $N=13$ and $d=2$.
 }
 \label{fig:QV_results}
\end{figure}

We present cross-platform fidelity results for 7- and 13-qubit QV circuits\cite{jurcevic2021demonstration}.
QV circuits have been studied extensively, both theoretically and experimentally\cite{jurcevic2021demonstration,cross2019validating,pino2020demonstration}, making them an ideal choice for the cross-platform comparison. An $N$-qubit QV circuit consists of $d$ layers : each layer contains a random permutation of the qubit labels, followed by random two-qubit gates among every other neighboring pair of qubits. Specifically, a QV circuit can be written as a unitary operation $V = \prod_{i=1}^d V^{(i)}$, where $ V^{(i)} = V^{i}_{\pi_i(N'-1), \pi_i(N')}\otimes \cdots \otimes V^{i}_{\pi_i(1), \pi_i(2)}$ and $N' = 2\lfloor N/2 \rfloor$. The operation $\pi(a)$ is a random permutation sampled from the permutation group $S_{N}$. The unitary operation $V^{i}_{a, b}$ is a random two-qubit gate acting on qubits $a$ and $b$ and sampled from $SU(4)$. The circuit diagram of an example QV circuit is shown in Fig.~\ref{fig:QV_results} a. In this experiment, we infer the fidelity for 7-qubit QV states with $d = 2$ and $d = 3$ and a 13-qubit QV state with $d = 2$.

Similar to the GHZ case, we first distribute the circuits, synthesize them into device-specific native gates, and allow optimizations/error-mitigation that satisfies the aforementioned state-preparation rule. 

On each platform, we append the circuit with $M_U = 500$ different $U$s sampled using the greedy method. Outcomes are measured in the computational basis for $M_S = 2000$ shots. The cross-platform fidelities for $d = 2$ and $d = 3$ are shown in Figs.~\ref{fig:QV_results} c,d. Our results verify that with only a fraction of the number of measurements required to perform full quantum state tomography, we can estimate the fidelities to sufficiently high precision to be able to see clear differences among them.

We also infer the cross-platform fidelity with a 13-qubit QV circuit with $d=2$. The results are shown in Fig.~\ref{fig:QV_results}b. Here we use $M_U=1000$ and $M_S=2000$, in contrast with the much larger $M_U=3^{13} = 1594323$ needed for full quantum state tomography. 

We find several interesting features by analyzing the cross-platform fidelity of 7-qubit QV results. First, we observe that the cross-platform fidelity drops significantly when the number of layers $d$ increases from $d= 2$ to $d = 3$ for the IBM quantum computers. The drop may be due to the restricted nearest-neighbor connectivity of superconducting quantum computers \cite{linke2017experimental}, requiring additional SWAP gates overhead for the execution of the permutation gates. In supplementary Information Sec. \ref{supp:swap}, we numerically evaluate the number of entangling gates as function of the number of layers $d$ with different connectivity graphs. We see that, according to IBM's native compiler QISKit (see Supp. sec. \ref{supp:hardware}) extra entangling gates are used to perform two-qubit gates for non-nearest-neighbor qubits on superconducting platforms, resulting in extra errors.

The cross-platform fidelity between IBM\_2 and IBM\_3 is higher than the cross-platform fidelity between either of them and the ion-trap systems (and classical simulation) as shown in Fig.~\ref{fig:QV_results}c.  This motivates us to study whether quantum states generated from different devices tend to be similar to each other if the underlying technology of the two devices is the same. Therefore, we perform a further analysis to investigate this phenomenon, which we refer to as intra-technology similarity.

\begin{figure}[h]
\centering
 \includegraphics[width=0.7\textwidth]{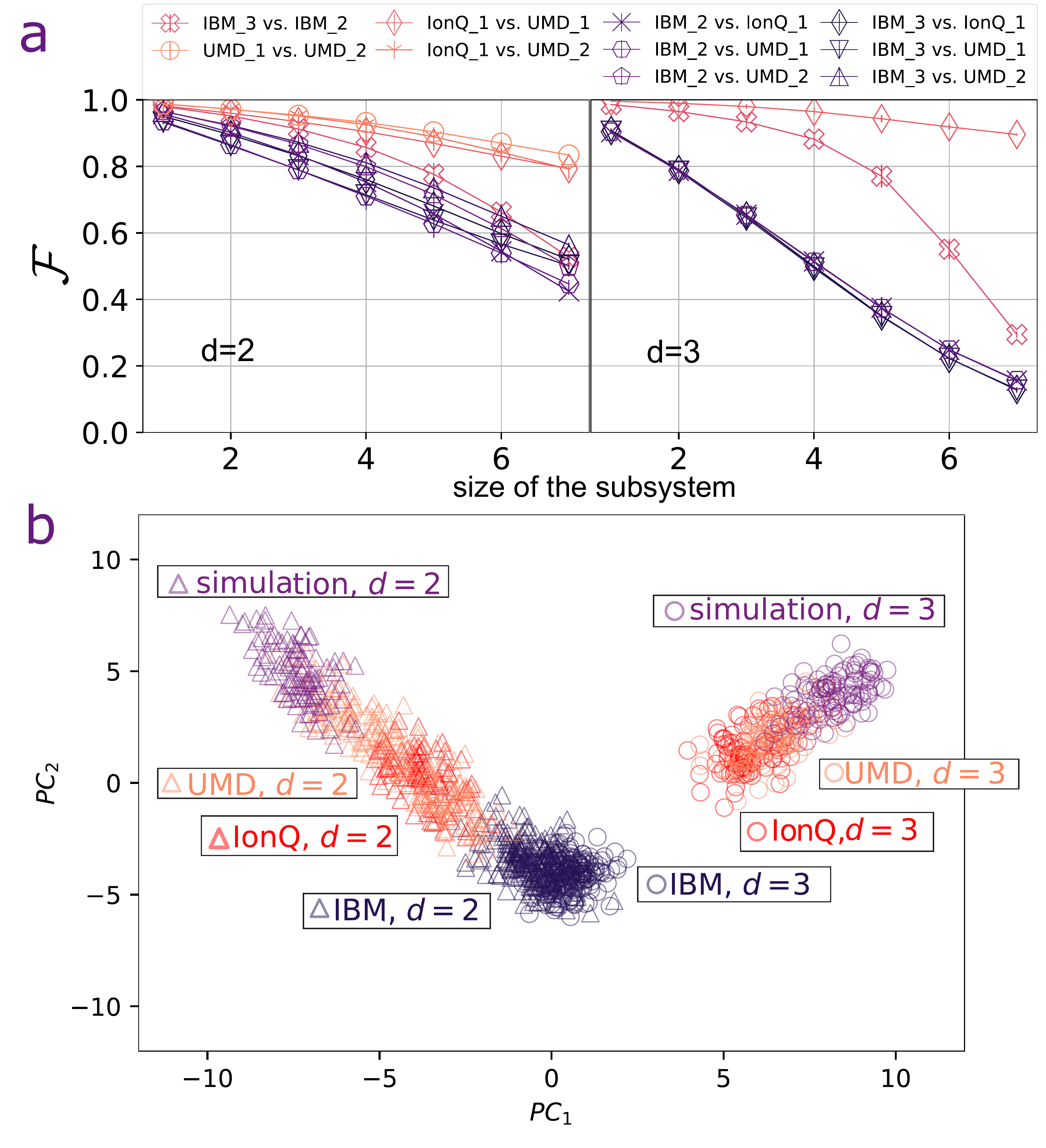}
 \caption{{\bf a} The cross-platform fidelity between subsystems prepared on different quantum computers. Left : 7-qubit quantum volume circuit of 2 layers. Right: 7-qubit quantum volume circuit of 3 layers. The mean and error for each subsystem size are calculated via bootstrap re-sampling.
{\bf b} The projection of randomized measurement dataset onto the first two principal axes, $PC_1$ and $PC_2$. Triangle marker is the 7-qubit quantum volume state with $d=2$. Circle marker is the 7-qubit quantum volume state with $d=3$. Magenta, orange ,and violet correspond to simulation, trapped-ion, and IBM systems respectively.}\label{fig:subsystem_fidelity}
\end{figure}

We first study the fidelity between subsystems of the 7-qubit QV states prepared on different quantum computers for both $d = 2$ and $d = 3$. The subsystem fidelity provides a scalable way to estimate the upper bound for the full system fidelity, since the cost of measuring all possible subsystem fidelities of a fixed subsystem size scales polynomially with the full system size. 
For a given subsystem, we use the same data collected for the full system, but trace out qubits not within the subsystem of interest. 
The results are presented in Fig. \ref{fig:subsystem_fidelity} a. We observe that the cross-platform fidelity between for all subsystem sizes from the same technology is higher for a given subsystem size. 

To further characterize the intra-technology similarity, we perform principal component analysis\cite{jolliffe2016principal} (PCA) on the randomized measurement data for the 7-qubit quantum volume states with $d=2$ and $d=3$ from all the platforms. PCA is commonly used to reduce the dimensionality of a dataset. It has been applied extensively in signal processing such as human face recognition and audio compression.  When implementing PCA, we project the dataset onto the first few principal components to obtain lower-dimensional data while preserving as much of the variation as possible. 

To prepare the data for PCA, we randomly sample $1000$ shots from the randomized measurement data out of $M_U\times M_S = 1,000,000$ for each platform. We identify the set of Pauli strings whose expectation values can be evaluated using the sample. We then evaluate the expectation value of these identified Pauli strings by taking the average over the samples, and repeat the sampling $N_{\mathrm{sample}} = 500$ times without replacement to make $N_{\mathrm{sample}}$ data points in the $4^N$ dimensional feature space. The feature vectors represent averaged classical shadow of the quantum state generated from the quantum computers \cite{huang2020predicting, huang2021provably}. We perform a rotation on the feature space and find the first two principal axes, which are the axes that show the two most significant variances on the dataset. Figure.~\ref{fig:subsystem_fidelity}b shows the projection of the $N_{\mathrm{sample}}$ data points to the first two principal axes. We observe that the first principal component separates the two quantum volume states, and the second principal component can distinguish the technology that generates the states. 
The clustering of the data from the same technology indicates that each technology may share similar noise characteristics that can be distinguished through the cross-platform fidelity and machine-learning techniques.

In this manuscript, we experimentally performed the cross-platform comparison of four quantum states allowing the characterization of the quantum states generated from different quantum computers with significantly fewer measurements than those required by full quantum state tomography. To expand our understanding of the intra-technology similarity, more quantum states should be studied. Our method could be extended to additional technological platforms such as Rydberg atoms and photonic quantum computers.
With the large volume of quantum data generated from the randomized measurement protocol, we have only begun to explore the possibilities that machine learning techniques can offer. We envision extensions of our method will be indispensable in quantitatively comparing near-term quantum computers, especially across different qubit technologies.






\section*{Acknowledgements} 
We acknowledge Andreas Elben, Behtash Babadi, Benoît Vermersch and Peter Zoller for helpful discussions. We acknowledge the use of IBM Quantum services; the views expressed are those of the authors, and do not reflect the official policy or position of IBM or the IBM Quantum team.
This work was supported by the ARO through the IARPA LogiQ program (11IARPA1008), the NSF STAQ Program (PHY-1818914), the AFOSR MURIs on Dissipation Engineering in Open Quantum Systems (FA9550-19-1-0399) and Quantum Interactive Protocols for Quantum Computation (FA9550-18-1-0161), the ARO MURI on Modular Quantum Circuits (W911NF1610349), and the U.S. Department of Energy Quantum Systems Accelerator (QSA) Research Center (DE-FOA-0002253). N.M.L. acknowledges support from the Maryland—Army-Research-Lab Quantum Partnership (W911NF1920181), the Office of Naval Research (N00014-20-1-2695), and the NSF Physics Frontier Center at JQI (PHY-1430094). A.M.G. is supported by a JQI Postdoctoral Fellowship.

\newpage

\renewcommand\thesection{}
\section*{SUPPLEMENTARY INFORMATION}

\renewcommand\theequation{S\arabic{equation}}    
\renewcommand\thetable{S\arabic{table}}    
\renewcommand\thefigure{S\arabic{figure}}    
\renewcommand\thesubsection{S\arabic{subsection}}
\renewcommand\thesection{S\arabic{section}}
\setcounter{section}{0}    
\setcounter{subsection}{0}    
\setcounter{equation}{0}  
\setcounter{table}{0}   
\setcounter{figure}{0}    

\section{Greedy method in the regime $M_S \gg 2^N$}
\label{supp:greedy_random}
The parameters $M_U$ and $M_S$ can be optimized through minimizing the statistical error with grid search \cite{crossplatfo, huang2020predicting} or using the perform importance sampling with partial information on the quantum state \cite{rath2021importance}. Both approaches require prior knowledge or simulation of the target state. Here, we devise a greedy method for sampling the unitary operation $U$ that reduces the statistical error without prior knowledge of the target state. The statistical error as a function of $M_U$ converges faster than uniformly sampling the unitary operation when the number of shots $M_S \gg 2^N$, where $N$ is the number of qubits. Therefore, the greedy method is particularly useful for 5- and 7-qubit experiments. In this section, we demonstrate the comparison between the greedy method and random method for 5-qubit GHZ state.

\begin{figure}[h]
\centering
 \includegraphics[width=0.7\textwidth]{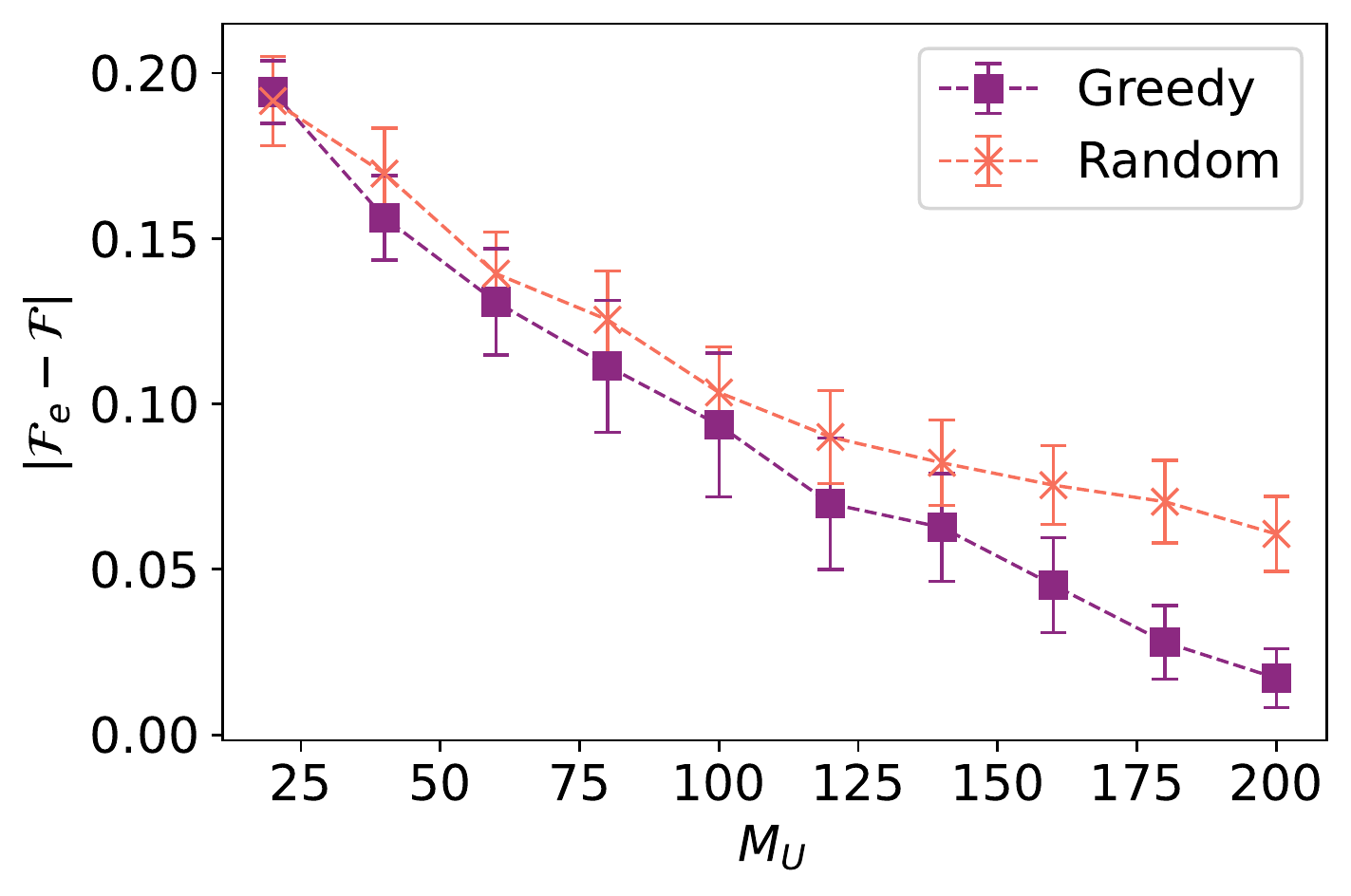}
 \caption{Comparison of error scaling for the fidelity of the GHZ states generated from UMD\_1 vs IBM\_1 with greedy or random sampling method for $M_U$.
 }
 \label{fig:Greedy_vs_Random}
\end{figure}

When performing the fidelity estimation using randomized measurement, there are two major source of errors, the shot noise error and the the error from the incomplete tomography. The shot noise error can be suppressed when the number of shots $M_S \gg 2^N$. In this section, we propose the greedy method for sampling the random unitary in order to mitigate the error from the incomplete tomography.
Instead of uniformly sampling the random unitary from a set of unitary operators $U$, we generate a sequence of unitary operators while maximizing the distance between each random unitary. Specifically, we define the distance between two unitary operators as $d(u_a, u_b) = \max_{\rho} ||u_a \rho u_a^\dagger-u_b \rho u_b^\dagger)||_1$. And we generate the $M_U$ unitary operators $\{u_i\}$, where $1 \leq i \leq M_U$ sequentially. For $i = 1$, we sample a unitary operator randomly from $V$. For $i > 1$, we search for a unitary operator $u_i$ that minimizes the cost function $C(u_i; u_1, \dots, u_{i-1}) = -\sum_{j = 1}^{i-1} d(u_i, u_j)$. In order to minimize the cost function efficiently, we randomly generate $N_{\rm sample}$ distinct unitary operators $u_{i,x}$, where $1 \leq x \leq N_{\rm sample}$ and we define $u_i = \min_{u_{i,x}}C(u_{i,x};u_1, \dots, u_{i-1})$. In practice, we find that $N_{\rm sample} = 200$ is enough to find the minimum for $N = 7$ and $V = Cl(2)^{\otimes N}$, where $Cl(2)$ is the single qubit Clifford group. The greedy method is summarized in Algorithm \ref{greedy}.

\begin{algorithm}[H]
  \caption{Greedy method for sampling random unitary}
  \label{greedy}
   {\bf Input : } Number of random unitary $M_U$, a set of unitary operator $S$\\
   {\bf Output :} $M_U$ random unitary operations for randomized measurement $\{u_i\}$, where $1\leq i \leq M_U$.
    
    ~~~1 : Sample $u_1$ randomly from $S$.
    
    ~~~2 : {\bf for} $i = 2$ {\bf to} $M_U$ {\bf do}
    
    ~~~3 : ~~~Find a unitary $u_i \in S$ to minimize the cost function $C(u_i; u_{1},\dots , u_{i-1})$.
    
    ~~~4 : {\bf end for}
    
    ~~~5 : {\bf return } $\{u_i\}$
    
\end{algorithm}

We compare the two different methods of sampling the random unitary $U$: the randomized sampling and the greedy method. Using these two methods, we evaluate the fidelity between the state prepared on the UMD\_1 system and that prepared on the IBM\_1 system, by sampling subset of various size $M_U$ from the full state tomography measurements. Fig.~\ref{fig:Greedy_vs_Random} shows the error of the fidelity estimation between UMD\_1 and IBM\_1 as function of $M_U$ for $M_S = 2000$. We see that the greedy method outperforms the random method in this regime.

\section{Full state tomography vs. randomized measurement for 5-qubit GHZ state}
\label{supp:5q-ghz}

Here, we compare the cross-platform fidelity obtained from full-state tomography and that from the randomized measurement on the 5-qubit  GHZ state prepared on different platforms.
We perform the full-state-tomography on a platform by measuring all the 243 independent 5-qubit Pauli operators. To do so, we first independently generate the 5-qubit GHZ state circuits on each platform, with all the optimizations that satisfy the application based criterion described in the main text. Then we append different single-qubit rotations to the circuit to create the 243 different circuits. Each of the circuits gives the projective measurement result of one of the 243 independent 5-qubit Pauli operators. We set $M_S=2000$ for all the platforms. 
For the randomized measurement, because a random Pauli basis measurement is equivalent to a randomize measurement with single qubit Clifford gate \cite{huang2020predicting}, we directly sample from the 243 Pauli basis measurements used for the full state tomography.

\begin{figure}[h]
\centering
 \includegraphics[width=0.7\textwidth]{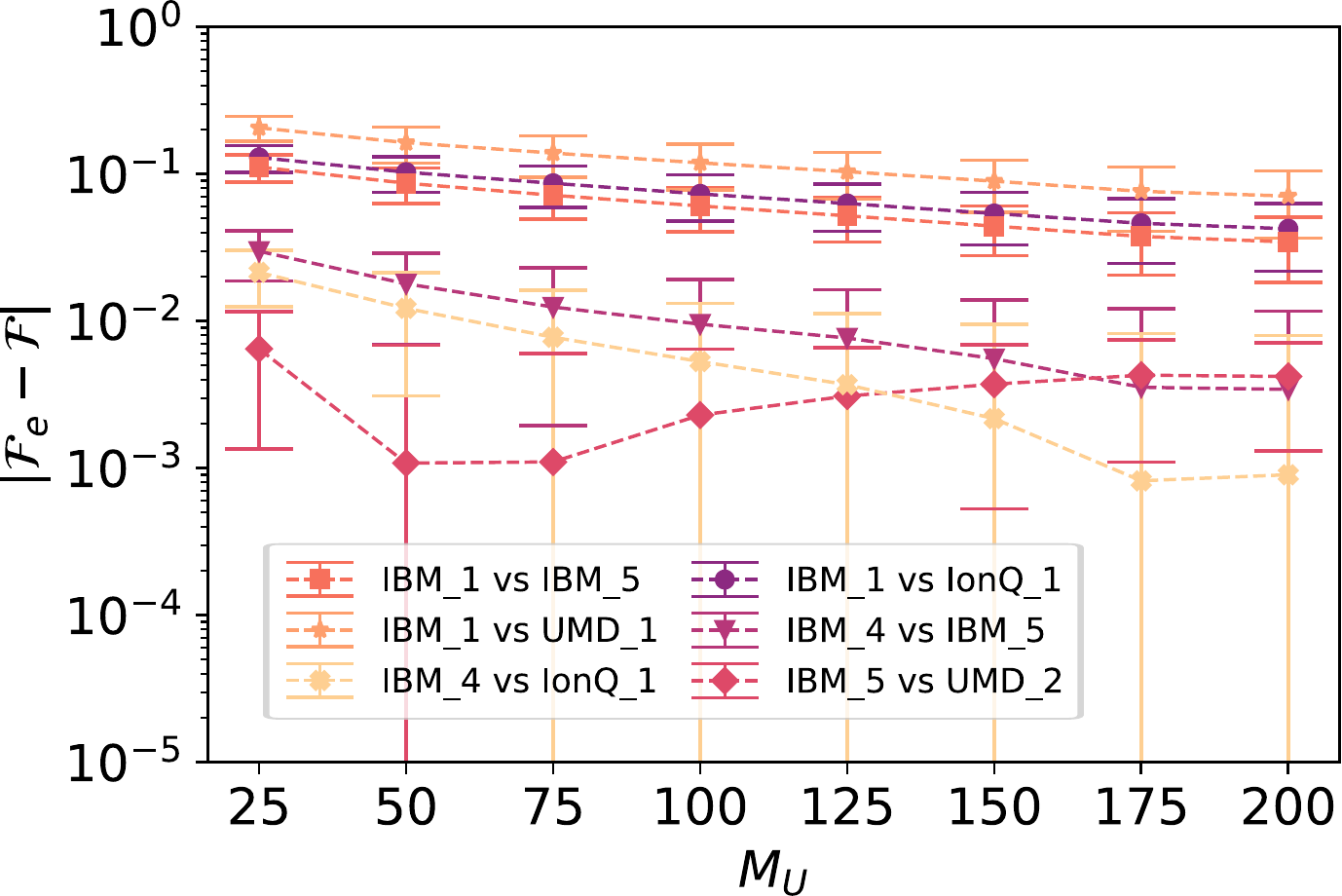}
 \caption{Fidelity error, $|\mathcal{F}_e-\mathcal{F}|$, for 5 randomly selected 5-qubit GHZ state cross-platform fidelities implemented on different platforms vs. number of randomized measurements $M_U$. The number of measurement is $M_S=2000$ for all cases.}
 \label{fig:GHZ_fidelity}
\end{figure}

We calculate the cross-platform fidelity between UMD\_1 and other platforms as function of the number of randomized measurements $M_U$. The fidelity error $|\mathcal{F}_e - \mathcal{F}|$ is defined as the difference between the fidelity estimated by the randomized measurement $\mathcal{F}_e$ and the fidelity calculated through full state tomography $\mathcal{F}$. The averaged error $|\mathcal{F}_e - \mathcal{F}|$ and the standard deviation are calculated through bootstrap resampling method \cite{efron1983leisurely}. The result (Fig.~\ref{fig:GHZ_fidelity}) shows that with only a fraction of the full state tomography measurements, one can estimate the cross-platform fidelity accurately. 

\section{SWAP overhead for quantum volume circuit}
\label{supp:swap}
Two-qubit gates on non-nearest-neighbor pairs are not directly available on superconducting quantum computers. To realize such non-nearest-neighbor two-qubit gates effectively, extra SWAP gates are necessary.  Each SWAP gate consists of three CNOT gates, which cause non-trivial degradation to the overall fidelity of a circuit.

\begin{figure}[htbp]
\centering
\includegraphics[width=0.8\textwidth]{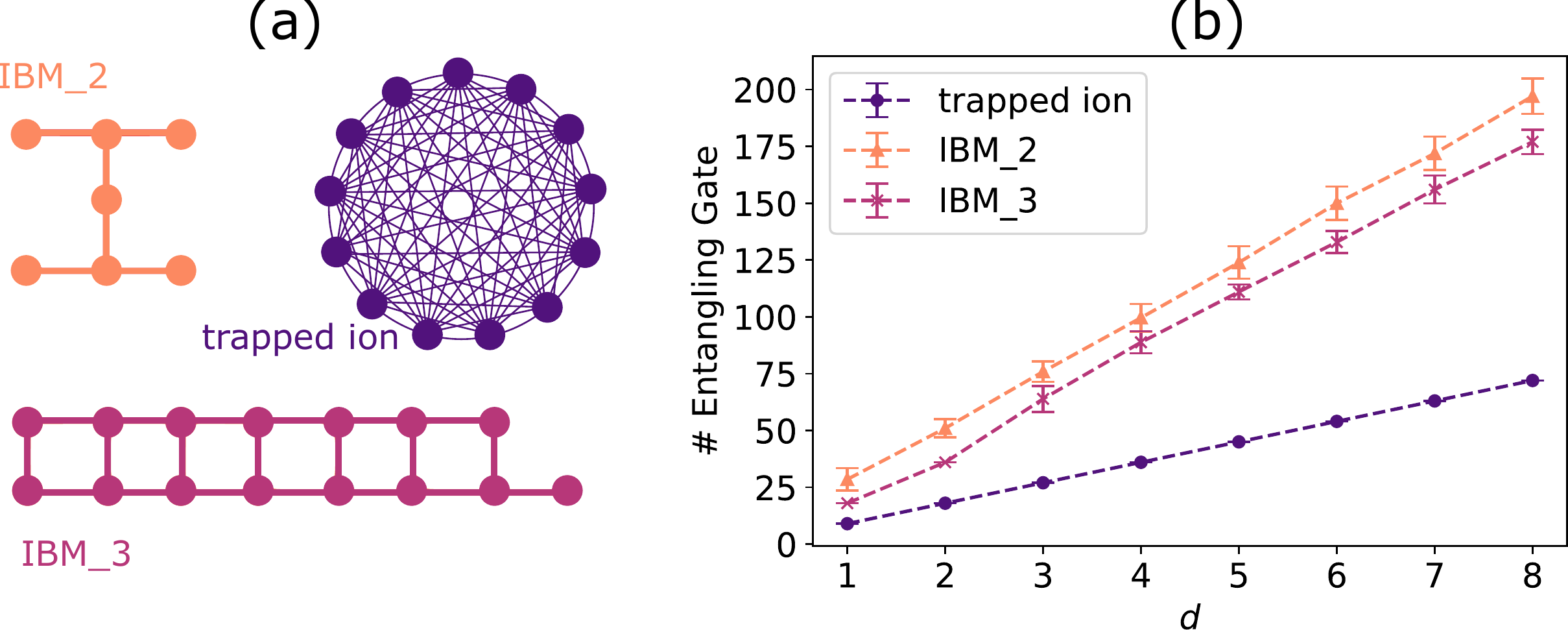}
\caption{(a) Connectivity graph of IBM\_2, IBM\_3, and trapped ion (UMD\_1 as an example) (b) Average number of two-qubit (entangling) gates needed to implement quantum volume circuits of layer $d$, on different quantum computers. The trapped ion quantum computers have the same all-to-all connectivity.}\label{fig:N_circuit_vs_N_layers}
\end{figure}

Optimizing the qubit routing can effectively decrease the number of involved non-nearest-neighbor two-qubit gates in evaluating the quantum volume circuits. But as the number of layers $d$ increases, the number of non-nearest-neighbor two-qubit gates needed increases. In fig. \ref{fig:N_circuit_vs_N_layers} we show the mean value of two-qubit gates needed to implement quantum volume circuits of $d$ layers on different platforms. 
As shown in the figure, the extra overhead grows linearly with $d$.

\section{Quantum systems}
\label{supp:hardware}

In this section we detail the quantum systems used in this study.


{\bf IBM Quantum Experience}

We use IBM Quantum Experience service to access several of their superconducting quantum computers. \cite{IBMquantum}
The ones used are \emph{ibmq\_belem} (IBM\_1), \emph{ibmq\_casablanca} (IBM\_2), \emph{ibmq\_melbourne} (IBM\_3), \emph{ibmq\_quito} (IBM\_4), and \emph{ibmq\_rome} (IBM\_5).
All the IBM systems use superconducting transmon qubits. The native gate sets are made of arbitrary single qubit rotations and nearest-neighbor two-qubit CNOT gates according to the connectivity graph.
The error of single-qubit gates in IBM systems ranges from $3.32\times 10^{-4}$ to $5.03\times 10^{-2}$, and the two-qubit errors range from $7.47\times 10^{-3}$ to $1.07\times 10^{-1}$.
Detailed specifications of each quantum device including qubit-connectivity diagram can be found on (\url{https://quantum-computing.ibm.com/}).  On this platform, the synthesis and circuit optimization are implemented using the QISKit open-source software \cite{cross2018ibm}.

{\bf TI\_EURIQA (UMD\_1)}

Error-corrected Universal Reconfigurable Ion-trap Quantum Archetype (EURIQA) is a trapped-ion quantum computer currently located at the University of Maryland. This quantum computer supports up to thirteen qubits in a single chain of fifteen trapped $^{171}{\rm Yb}^+$ ions in a microfabricated chip trap~\cite{Maunz2016}. The system achieves native single-qubit gate fidelities of 99.96\% and two-qubit XX gate fidelities of 98.5-99.3\%\cite{egan2020fault}. On this platform, we compile the circuits to its native gate set through KAK decomposition. We optimize the qubit assignment through exhaustive search to minimize the anticipated noise of entangling gates. No SPAM correction was applied in post-processing.

{\bf TI\_UMD (UMD\_2)}

The second trapped-ion quantum computer system at Maryland is part of the TIQC (Trapped Ion Quantum Computation) team. This quantum computer supports up to nine qubits made of a single chain of ${}^{171}\text{Yb}^+$ ions trapped in a linear Paul trap with blade electrodes \cite{debnath2016demonstration}. Typical single- and two-qubit gate fidelities are $99.5(2)\%$ and $98-99\%$, respectively. On this platform, we compile the quantum volume to its native gate set through KAK decomposition. We apply SPAM correction to mitigate the detection noise assuming that the preparation noise is negligible.

{\bf IonQ (IonQ\_1 and IonQ\_2)}

The commercial trapped-ion quantum systems used by IonQ contain eleven fully connected qubits in a single chain of ${}^{171}\text{Yb}^+$ ions trapped in a linear Paul trap with surface electrodes \cite{debnath2016demonstration}.
The single-qubit fidelities are $99.7\%$ for both systems at the time of measurement, while two-qubit fidelities are $95-96\%$ and $96-97\%$ for IonQ\_1 and IonQ\_2 respectively. On this platform, we apply the technique describe in Ref. \cite{nam2018automated} to optimize the circuit. Quantum volume circuits were decomposed in terms of partially entangling MS gates. No SPAM correction was applied in post-processing.




\bibliographystyle{naturemag}
\bibliography{reference}

\end{document}